\documentclass[a4,aps,twocolumn,prl,amsmath,showpacs,amssymb,superscriptaddress]{revtex4}
\usepackage{graphicx}% Include figure files
\usepackage{dcolumn}% Align table columns on decimal point
\usepackage{bm}% bold math

\begin{document}

\title{Mode-Locking in a Free-Electron Laser Amplifier}

\author{N. R. Thompson}
\email{n.r.thompson@dl.ac.uk} \affiliation{University of Strathclyde
(SUPA), Glasgow G4 0NG, UK} \affiliation{ASTeC, Daresbury
Laboratory, Warrington WA4 4AD, UK.}
\author{B. W. J. M$^{\mathrm{c}}$Neil}
\email{b.w.j.mcneil@strath.ac.uk} \affiliation{University of
Strathclyde (SUPA), Glasgow G4 0NG, UK}

\date{\today}

\begin{abstract}
A technique is proposed to generate attosecond pulse trains of
radiation from a Free-Electron Laser amplifier. The optics-free
technique synthesises a comb of longitudinal modes by applying a
series of spatio-temporal shifts between the  co-propagating
radiation and electron bunch in the FEL. The modes may be
phase-locked by modulating the electron beam energy at the mode
spacing frequency. Three-dimensional simulations demonstrate the
generation of a train of 400~as pulses at giga-watt power levels
evenly spaced by 2.5~fs at a wavelength of 124~$\mathrm{\AA}$. In
the X-ray at wavelength 1.5~$\mathrm{\AA}$, trains of 23~as pulses
evenly spaced by 150~as and of peak power up to 6 GW are predicted.
\end{abstract}

\pacs{41.60.Cr 42.55.Vc 42.60.Fc }

\maketitle

%\section{Introduction}
Attosecond pulses of high power, short wavelength, electromagnetic
radiation would enable observation and possible control of very fast
phenomena at the atomic timescale~\cite{attosecondscience}. A new
source of high power X-ray pulses is the high gain Free-Electron
Laser (FEL)~\cite{fels}. However, in the Self Amplified Spontaneous
Emission mode of operation~\cite{kondratenko,bnp}, the radiation has limited temporal coherence~\cite{bonifacioPRL} with
widths approximately that of the electron bunch source, typically
many tens of femtoseconds. Several techniques have already been
identified that may shorten this timescale into the
attosecond regime. Selection of a short harmonic spike in a
multiple undulator harmonic cascade may be one
option~\cite{saldin2002}. Other techniques pre-modulate the electron
bunch energy with an optical laser before the bunch enters the
radiator undulator. The resonant FEL wavelength is
correlated to this energy modulation and it may be possible to
selectively filter~\cite{saldin2004_1} or
amplify~\cite{saldin2004_2} a narrow wavelength band to generate
short pulses with widths of a fraction of the modulation period.
Other techniques rely upon similar electron bunch energy modulation
methods~\cite{fawley,zohlents-penn2005}. Another approach, called
\hbox{E-SASE}~\cite{esase1,esase2}, uses an optical laser to
modulate the electron bunch energy at an intermediate acceleration
stage. Regions of enhanced current are created that subsequently
generate short pulses in a final radiator undulator. A conceptually
simpler method has also been suggested~\cite{emma} that `spoils' all
but a short region of the electron bunch. Only this region
subsequently lases in the FEL to generate a short radiation pulse.
More recently, a technique using an FEL with a negative undulator
taper and a pre-modulated bunch energy has been
proposed~\cite{saldin2006}. The above techniques typically predict
pulse widths $\agt 100$~as at a target wavelength of $\sim
1.5$~$\mathrm{\AA}$. 

Here, a technique is presented that may shorten the duration of
X-ray pulses generated by SASE FELs to less than the atomic unit of
time (24~as). By applying concepts from mode-locked cavity
lasers~\cite{siegman}, we predict generation of a train of multi-GW
peak power pulses, at wavelength 1.5~$\mathrm{\AA}$,  of width
$\approx 23$~as with 150~as separation, and  a power contrast ratio
of approximately sixty. Such pulses may have sufficient power,
spatial and temporal resolution to offer a new scientific tool for
observing the dynamics of atomic-scale phenomena.

%\begin{figure*}[htb]
\begin{figure}
\includegraphics[width=80mm]{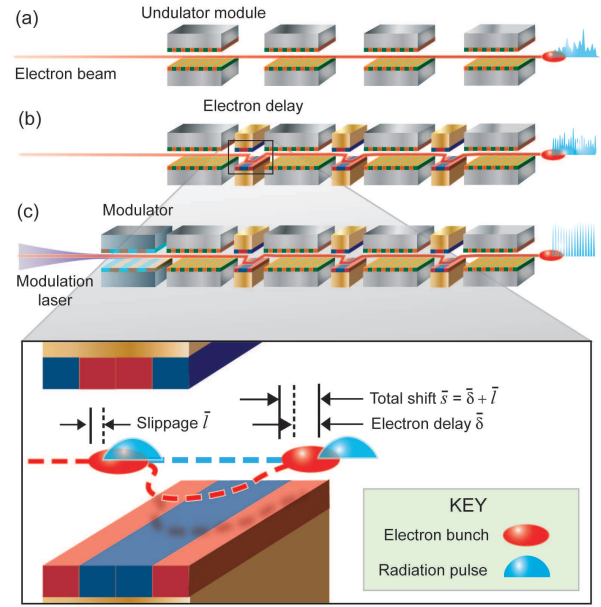}
\caption{\label{fig1}Schematic of three regimes of FEL interaction:
(a) SASE regime (b) Mode-coupled SASE regime and (c) Mode-locked
SASE regime. The inset shows detail of the electron delay.}

\end{figure}

In a self-amplified spontaneous emission (SASE) FEL, shown
schematically in Fig.~\ref{fig1}~(a), a relativistic electron beam
propagates at mean velocity $v_z < c$ along an undulator and
exponentially amplifies the initial spontaneous emission via the FEL
instability~\cite{kondratenko, bnp}. Radiation emission is at the
resonant FEL wavelength
$\lambda_r=\lambda_u(1+\bar{a}_u^2)/2\gamma_0^2$ where $\lambda_u$
is the undulator period, $\bar{a}_u$ is the rms undulator parameter
and $\gamma_0$ is the mean electron bunch energy in units of the
electron rest mass energy. For exponential growth the undulator must
be longer than the gain length of the FEL interaction
$l_g=\lambda_u/4\pi\rho$~\cite{bnp}, where $\rho$ is the
dimensionless FEL parameter, a measure of the coupling strength
between electron beam and radiation, having typical values $\rho\sim
10^{-3}$ to $10^{-2}$.

The spatio-temporal range of the SASE interaction is
different from a conventional laser: in the latter a radiation
wavefront interacts with the full length of the gain medium, whereas
in a FEL a wavefront may propagate through only a fraction of a
typical electron bunch as $v_z\alt c$. Autonomous regions of the
radiation/electron interaction evolve from noise and are
uncorrelated in phase. The temporal output power is noisy,
comprising phase-uncorrelated pulses with separation $\lesssim 2\pi
l_c$, where the cooperation length
$l_c=\lambda_r/4\pi\rho$~\cite{bmp} is the relative propagation
between a radiation wavefront and electron bunch in one gain length
and defines the intrinsic coherence length of the radiation
$l_{\mathrm coh}\approx l_c$. The SASE spectrum at saturation is
similarly noisy comprising irregular spikes within rms bandwidth
$\sigma_\lambda/\lambda_r \approx \rho$.

Here, techniques similar to those used in mode-locked broad
bandwidth cavity lasers are used to improve the quality of the FEL
output over SASE. The method generates a set of
axial radiation modes, analogous to the axial cavity modes in
conventional laser oscillators~\cite{siegman} which may be
(actively or passively) phase locked by introducing a modulation
with frequency $\Delta \omega_s$ equal to the mode spacing. Each
mode acquires sidebands that overlap neighbouring modes. The
mode phases lock to generate temporal pulses spaced at, and with
widths significantly shorter than, the cavity transit time
\cite{siegman}.

The axial modes in the SASE FEL amplifier are synthesised by
applying a series of spatio-temporal shifts between the radiation
and co-propagating electron bunch. These shifts are achieved by
repeatedly delaying the electron bunch using magnetic chicanes
between undulator modules, as shown schematically in
Fig.~\ref{fig1}~(b). This `slippage' of the electron bunch behind
the radiation is in addition to the slippage occurring within the
undulator modules and increases the spatio-temporal coupling range
over that of the simple SASE case, increasing both the cooperation
and coherence lengths of the interaction. For equal,
periodic delays, the radiation spectrum develops discrete
frequencies similar to the axial modes of a conventional laser.
These modes are intrinsically coupled via their collective
interaction with the electron bunch over regions of a coherence
length $l_{coh}$ within which a series of regularly spaced radiation
spikes of approximately uniform width $\ll l_{coh}$ may evolve. Such
a system, illustrated in Fig.~\ref{fig1}~(b), is termed
`mode-coupled'. By introducing an interaction modulation at the mode
 frequency spacing $\Delta \omega_s$, for example a coherent
modulation of the input electron bunch energy shown schematically in
Fig.~\ref{fig1}~(c), the modes may become phase locked to give a
temporal train of equally spaced, short, high-power pulses phase
correlated over a distance~$\gg l_{coh}$. The system is then termed
`mode-locked'.

%\section{Analysis}
The effect of the spatio-temporal shifts on the radiation spectrum
may be seen by considering the radiation emitted from a small,
constant electron source term. In the universal scaling
of~\cite{bnp,bmp} the one-dimensional wave equation describing the
field evolution may be written:
\begin{equation}\label{waveequation}
\frac{\partial A(\bar{z},\bar{z}_1)}{\partial
\bar{z}}+\frac{\partial A(\bar{z},\bar{z}_1)}{\partial
\bar{z}_1}=b_0(\bar{z}_1),
\end{equation}
where $A$ is the scaled electric field, $\bar{z}$ is the interaction
length along the undulator in units of $l_g$, $\bar{z}_1$ is the
scaled electron bunch coordinate in units of $l_c$ and the small
initial electron source term $b_0\ll 1$.
%Normally $b_0$ would also be a function of $\bar{z}$, however, as any initial bunching would decay as $|b_0(\bar{z},\bar{z}_1)|\approx |b_0(\bar{z}_1)|\exp(-\bar{z}^2\bar{\sigma}_\gamma^2/2)$~\cite{bdecay}, where $\bar{\sigma}_\gamma$ is the scaled effective energy spread of the beam~\cite{emittance}, (\ref{waveequation}) will be valid in the limit $\bar{z} \ll 1/\bar{\sigma}_\gamma$.
A solution to equation~(\ref{waveequation}) for a single
undulator/chicane  module may be obtained using a Fourier transform
with respect to $\bar{z}_1$ with no initial radiation
$\tilde{A}(0,\bar{\omega})= 0$. For a series of $N$
undulator/chicane  modules a solution may be obtained by application
of the Fourier transform time-shifting  relation to yield:
%\begin{equation}\label{FT}
%\frac{d\tilde{A}(\bar{z},\omega)}{d \bar{z}}+i\omega
%\tilde{A}(\bar{z},\omega)=\tilde{b}_0(\omega).
%\end{equation}
\begin{equation}\label{shiftspectrum2}
\tilde{A}(\bar{L},\bar{\omega})=
\tilde{b}_0 \;\bar{l}\;\mathrm{sinc}(\bar{\omega}\bar{l}/2)\;%\sum_{n=1}^{N} e^{i(n-1)\omega \bar{s}}
\frac{e^{iN\bar{\omega}\bar{s}}-1}{e^{i\bar{\omega}\bar{s}}-1}\;e^{-i\bar{\omega}
\bar{l}/2},
\end{equation}
where the scaled interaction length $\bar{L}=N\bar{l}$, the total slippage per module in units of $\bar{z}_1$ is
$\bar{s}=\bar{l}+\bar{\delta}$, $\bar{l}$ is the slippage due to the
undulator,  $\bar{\delta}$ is the the slippage due to the chicane,
the transform variable
$\bar{\omega}=(\omega-\omega_r)/2\rho\,\omega_r$ and $\omega_r$ is
the resonant FEL frequency. The expression~(\ref{shiftspectrum2}) is
identical in form to the result of a simple analysis that gives the
axial mode content of a conventional cavity laser~\cite{siegman}.

It is useful to define the slippage enhancement factor
$S_e=\bar{s}/\bar{l}$. The extra slippage increases both $l_c$ and the $l_{coh}$ by
the factor $S_e$. For $S_e=1$ there are no
chicane sections ($\bar{\delta}=0$) and~(\ref{shiftspectrum2})
reduces to the spontaneous emission spectrum for an undulator of
scaled length $\bar{L}=N\bar{l}$. For $S_e>1$ the
sinc-function envelope (the single undulator module spectrum) is
modulated by  a `frequency comb' centred at the scaled resonant
frequency $\bar{\omega}=0$, with mode separation
$\Delta\bar{\omega}=2\pi/\bar{s}$ corresponding to a
$\Delta\omega=2\pi/T_s$ where $T_s=s/c=\bar{s}l_c/c$ is the time
taken for radiation to travel the slippage length. In effect, the
spectrum is that of a ring cavity of length $s$. Each sideband mode
has width $\delta\bar{\omega}\propto (S_eN)^{-1}$.
%Examples of the spectrum (\ref{shiftspectrum2}) are shown in Fig.~\ref{spectrum} for $S_e=2$ and $4$.
The $k$th sideband at frequencies $\bar{\omega}_k=\pm
k\Delta\bar{\omega}$ will fall under the central peak of the
sinc-function when $S_e>k$. For the case $N=2$, equation
(\ref{shiftspectrum2}) reduces to the spectrum of an Optical
Klystron (OK) and agrees well with measured OK
spectra~\cite{SuperACO,elettra}.

\begin{figure}
\includegraphics[width=85mm]{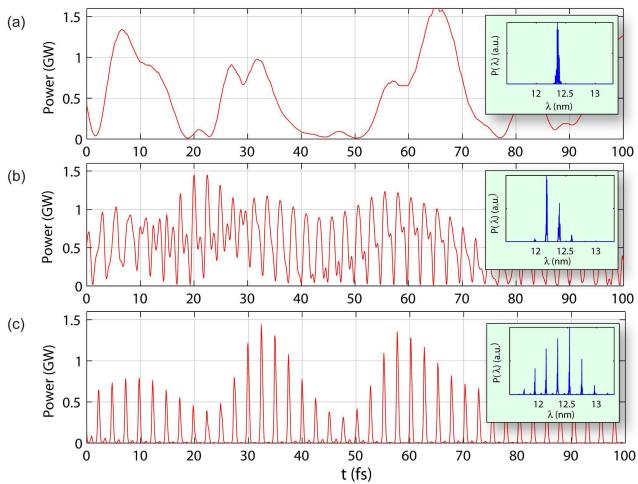}
\caption{\label{fig2}The radiation power as a function of time $t$
with radiation power spectrum as a function of wavelength in nm
(inset) for the three cases of (a) SASE, (b) mode-coupled and (c)
mode-locked.}
\end{figure}

The delay chicanes may also introduce longitudinal energy dispersion
parameterized via $D=R_{56}/2l_c$~\cite{chicane,bonifacio2}, which
for $D>0$  enhances the FEL electron gain process so reducing $l_g$
and $l_c$ by a nominal factor $G_e\leq 1$.
Such enhanced gain is the principle behind the Distributed Optical
Klystron (DOK) FEL amplifier~\cite{litvinenko,DOK,ding}. The
combined effects of the enhanced slippage ($S_e\geq 1$) and
gain length reduction  ($G_e<1$) give a modified
cooperation length $\hat{l}_c=S_e G_e l_c$.

\begin{figure}
\includegraphics[width=85mm]{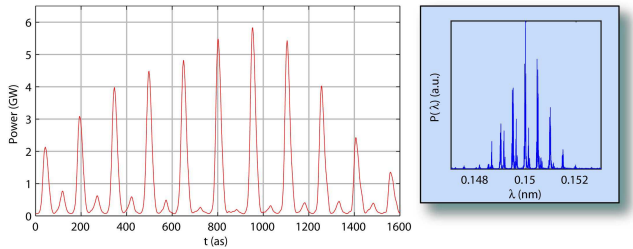}
\caption{\label{fig3}A mode-locked $1.5\mathrm{\AA}$ X-ray FEL
example: the radiation power as a function of time $t$ with
radiation power spectrum as a function of wavelength in nm.}
\end{figure}

As illustrations of the technique two systems were simulated using
the 3-D code Genesis~1.3~\cite{genesis}, used in the design of major
facilities such as LCLS~\cite{lcls} and XFEL~\cite{xfel}. Parameters
are given in Table~\ref{tab:XUVAndXRayParameters}. The first system
is a typical XUV FEL design for operation at $124\mathrm{\AA}$. The
radiation power output close to saturation over a 100fs sample for
the three cases of SASE, mode-coupling and mode locking of
Fig.~\ref{fig1}(a..c) respectively are shown in Fig.~\ref{fig2}.
Insets plot the power spectral density as a function of radiation
wavelength. For the SASE case the output is seen to be noisy,
comprising a series of irregularly spaced pulses with mean
separation approximately $2\pi l_{c}$, corresponding to a temporal
duration $2\pi\tau_{coh}=\lambda_r/2c\rho\approx 8$fs, with a
spectrum centred at the resonant wavelength of fractional
width~$\approx 2\rho$, results typical of the SASE
regime~\cite{bonifacioPRL}.
\begin{table}
    \centering
        \begin{tabular}{|l |c| c|}
        \hline
         & XUV & X-ray\\
        \hline
         Bunch energy E(GeV)& 0.75 & 14.3\\
        \hline
         Bunch peak current I(kA)& 3 & 3.4\\
        \hline
        % Bunch charge (nC)& 1 & 1\\
        %\hline
         Normalised emittance $\epsilon_n$(mm-mrad)& 2 & 1.2\\
        \hline
         RMS fractional energy spread $\sigma_\gamma/\gamma_0$& $10^{-4}$ & $8\times 10^{-5}$\\
        \hline
         Undulator period $\lambda_u$(cm) & 3.1 & 3\\
        \hline
         Resonant wavelength $\lambda_r$(\AA) & 123 & 1.5\\
        \hline
         Undulator module length (units $l/\lambda_u$)& 12 & 72\\
        \hline
        % Average beta function $\beta$(m) & 9.6 & 37.2\\
        %\hline
         FEL parameter $\rho$ & $2.5\times 10^{-3}$ & $5\times 10^{-4}$\\
        \hline
          Chicane delay $N_c=\delta/\lambda_r$ & 48 & 228\\
        \hline
          Modulation period (units of $\lambda_r$)&61 & $303$\\
        \hline
          Modulation amplitude (MeV)&5.8 & 14.3\\
        \hline
        Slippage enhancement $S_e$ & 5 & 5\\
        \hline
        \end{tabular}
    \caption{XUV and X-ray simulation parameters.}
    \label{tab:XUVAndXRayParameters}
\end{table}
In Fig.~\ref{fig2}(b), the mode-coupled case, chicanes are
introduced to give delay $\delta=48\lambda_r$, with total slippage
due to undulator and delay section $s=60\lambda_r$, a total slippage
time of $T_s\approx 2.48$fs, slippage enhancement $S_e=5$ and gain
length reduction factor $G_e\approx 0.65$. The spectrum has sideband
modes with separation $\Delta\omega_s=2\pi/T_s$, giving wavelength
separation $\Delta\lambda\approx \lambda_r^2/s\approx0.21$nm. The
power comprises a series of spikes with  width $\tau_p\simeq 1 $fs
FWHM and separation $T_s\approx 2.48$fs. The mode spiking can only
be phase-matched over $\tau_{coh} = \hat{l}_c/c\approx 4.3$fs. This
results in the phase drifting of the spikes observed over the spike
envelope.

% that is modulated with period determined by $2\pi
%\tau_{coh}\approx 27$fs.

When the beam energy is modulated at the mode separation frequency
$\Delta\omega_s$, as shown schematically in Fig.~\ref{fig1}(c), the
radiation modes may phase lock over the entire bunch. A modulation
amplitude of 5.8~MeV was induced by seeding a 10 period modulator
undulator, $\lambda_u=75$mm, by a 230MW 755nm laser. From
Fig.~\ref{fig2}(c) it is seen that mode-locking occurs, dramatically
improving the temporal pulse structure over the SASE and
mode-coupled cases of Fig.~\ref{fig2}(a,b) to give a train of
pulses, evenly spaced by $T_s\approx 2.48$fs, of constant width
$\tau_p \approx 400$as and peak power $\lesssim 1.4$GW. The contrast
ratio is $> 100$. The only remnant of the SASE noise is the slowly
varying envelope of mean period $2\pi \tau_{coh}\approx 27$fs.

%Compared to the SASE case, mode-locking has reduced the pulse width by a factor of $\approx$ 25 and replaced the temporal randomness of the output with a train of evenly spaced spikes.
The second system demonstrates the FEL mode-locking technique scaled
to shorter wavelengths. Fig.~\ref{fig3} plots the simulated
saturated output power in the X-ray at $1.5$\AA~for $S_e=5$ and
other parameters similar to LCLS~\cite{lcls}. The pulse train
consists of $\approx 23$~as pulses separated by $T_s\approx 150$~as
with peak powers up to 6~GW and a contrast ratio $\approx 60$.

The output pulse width of a homogeneously broadened mode-locked
cavity laser is given by~\cite{siegman} $\label{spikewidth}
\tau_p\approx 0.5/\sqrt{N_0}f_m$ where $N_0$ is the number of
oscillating modes in the cavity and $f_m$ the modulation frequency.
Following the discussion of equation~(\ref{shiftspectrum2}), the
number of modes under the central peak of the spectrum is $N_0=2
S_e-1=9$, in good agreement with Fig.~\ref{fig2} and Fig.~\ref{fig3}
where $N_0\approx 8$ in the XUV and $N_0\approx 9$ in the X-ray.
The previous expression then gives pulse lengths $\tau_p\approx
440$as in the XUV and $\tau_p\approx 25$as in the X-ray, in good
agreement with the simulated values of $\tau_p\approx 400$as and
$\tau_p\approx 23$as respectively, further strengthening the analogy
with the results of conventional mode-locked cavity lasers and
indicating that the modes are phase-locked.

We stress that the mode-locking technique described in this letter
does not rely upon the dispersive properties of the chicanes on the
electrons -- what is important is the relative electron
bunch/radiation delay that creates the longitudinal modes. The
technique works equally well for isochronous chicanes with $D=0$.
Contrarily, the optical klystron effect relies only upon the chicane
dispersion, requiring $D>0$, and does not rely upon any relative
electron bunch/radiation delay. Furthermore,  the role of the
initial beam energy modulation in the mode-locking technique is  to
allow coupling and  phase-locking of the synthesised longitudinal
modes, whereas in the techniques
of~\cite{saldin2002,saldin2004_1,saldin2004_2,fawley,esase1,esase2,emma,zohlents-penn2005,saldin2006},
the beam energy modulation is primarily to allow either dispersive
electron beam manipulation or to introduce energy chirp effects
before radiation generation in the final FEL undulator sections.

A criterion for operation relates the electron beam energy spread to
the dispersive strength of the chicane~\cite{bonifacio2}: $
\bar{\sigma}_\gamma \equiv \sigma_\gamma/\rho\gamma_0\lesssim 1/D$.
Approximating $R_{56}\simeq10\delta/3$ gives $
\sigma_\gamma/\gamma_0\lesssim 1/20 N_c$ where $N_c =
\delta/\lambda_r$. The effect of magnet stability of the chicanes
upon the electron delay $\delta$ must also be considered. For a
4-dipole chicane with equal magnet and drift lengths, $L$, the delay
may be written $\label{delta} \delta=L \phi^2$, where $\phi$ is the
small deflection angle. In a single dipole of field strength $B$,
$\phi=eBL/\gamma_0mc$ so that $\Delta\delta=2\delta\Delta B/B $. Applying the
requirement for sub-wavelength phase matching $\Delta\delta <
\lambda_r$, and assuming that for $N$ independently powered chicanes
the errors add as in a random walk, the tolerance on the magnetic
field is then $\Delta B/B< 1/2N_c\sqrt{N}$. The effect of
shot-to-shot electron bunch energy fluctuation is also considered.
From the chicane delay $\delta=L \phi^2 $, and using the FEL
resonance condition, the delay may be expressed as
$\delta/\lambda_r=2e^2B^2L^3/m^2c^2\lambda_u(1+\bar{a}_u^2)$ which is
independent of the beam energy. Thus, beam energy fluctuations
should not affect the axial mode structure as the chicanes always
delay by the same number of resonant wavelengths. The numerical
values predicted by these tolerance criteria for the XUV (and X-ray) systems are as follows: $\sigma_\gamma/\gamma_0$ $<$ $1\times 10^{-3}$ ($2\times 10^{-4}$) and $\Delta B/B$ $<$ $2 \times 10^{-3}$ ($4 \times 10^{-4}$) demonstrating that both systems are feasible with current technology.

The introduction of mode-locking in conventional cavity lasers
dramatically increased the range of their scientific application. It has been shown here that the use of similar techniques in FELs yields a similar leap in capability for what are already recognised
as important tools for scientific investigation. Further advances may well be possible: alternative methods of
locking the coupled modes or generating more exotic modal structures
may enable pulse shaping in the attosecond domain; variation of the
pulse structure may be achievable by using different
chicane delays and tapered undulator sections.

\acknowledgments The authors would like to thank Steve Jamison and
Brian Sheehy for helpful discussions.

\end{document}